\documentclass[english,12pt]{article}
\usepackage{array}
\usepackage{graphicx}
\usepackage{amssymb}
\usepackage{amsmath}
\usepackage{multirow}
\usepackage{prettyref}
\usepackage{comment}
\usepackage[usenames, dvipsnames]{color}
\usepackage{babel}
\usepackage{units}
\usepackage[latin1]{inputenc}
\usepackage{amsfonts}
\usepackage{amssymb}
\usepackage{float}
\usepackage{cancel}
\usepackage{dsfont}
\usepackage{cite}
\def\@fmsl@sh#1#2#3{\m@th\ooalign{$\hfil#1\mkern#2/\hfil$\crcr$#1#3$}}
 \def\eq#1\en{\begin{equation}#1\end{equation}}
\def\s[#1,#2]{[#1\stackrel{\star}{,}#2]}
\def\sx[#1,#2]{[#1\stackrel{\star_{x}}{,}#2]}

\newcommand{\nc}{\newcommand}
\nc{\beq}{\begin{equation}}
\nc{\eeq}{\end{equation}}
\nc{\beqa}{\begin{eqnarray}}
\nc{\eeqa}{\end{eqnarray}}

\def\bc{\begin{center}}
\def\ec{\end{center}}

\def\to{\rightarrow}

\def\gsim{\mathrel{\mathpalette\atversim>}}

\def\bc{\begin{center}}
\def\ec{\end{center}}

\def\gsim{\mathrel{\rlap{\lower4pt\hbox{\hskip1pt$\sim$}}

    \raise1pt\hbox{$>$}}}       

\def\gsim{\mathrel{\rlap{\lower4pt\hbox{\hskip1pt$\sim$}}
    \raise1pt\hbox{$>$}}}       

\begin{document}
\makeatletter
\def\fmslash{\@ifnextchar[{\fmsl@sh}{\fmsl@sh[0mu]}}
\def\fmsl@sh[#1]#2{%
  \mathchoice
    {\@fmsl@sh\displaystyle{#1}{#2}}%
    {\@fmsl@sh\textstyle{#1}{#2}}%
    {\@fmsl@sh\scriptstyle{#1}{#2}}%
    {\@fmsl@sh\scriptscriptstyle{#1}{#2}}}
\def\@fmsl@sh#1#2#3{\m@th\ooalign{$\hfil#1\mkern#2/\hfil$\crcr$#1#3$}}
\makeatother

\thispagestyle{empty}
\begin{titlepage}
\boldmath
\begin{center}
  \Large {\bf Quantum Corrections to Schwarzschild Black Hole}
    \end{center}
\unboldmath
\vspace{0.2cm}
\begin{center}
{  {\large Xavier Calmet}\footnote{x.calmet@sussex.ac.uk} {\large and} 
{\large Basem Kamal El-Menoufi}\footnote{b.elmenoufi@sussex.ac.uk} 
}
\end{center}
\begin{center}
{\sl Department of Physics $\&$ Astronomy, 
University of Sussex, Falmer, Brighton, BN1 9QH, United Kingdom}
\end{center}
\vspace{5cm}
\begin{abstract}
\noindent
Using effective field theory techniques, we compute quantum corrections to spherically symmetric solutions of Einstein's gravity and focus in particular on the Schwarzschild black hole. Quantum modifications are covariantly encoded in a non-local effective action. We work to quadratic order in curvatures simultaneously taking local and non-local corrections into account. Looking for solutions perturbatively close to that of classical general relativity, we find that an eternal Schwarzschild black hole remains a solution and receives no quantum corrections up to this order in the curvature expansion. In contrast, the field of a massive star receives corrections which are fully determined by the effective field theory.
\end{abstract}  
\end{titlepage}

\section{Introduction}

With the recent discovery of gravitational waves, we have entered a new era in astrophysics in which we will be probing black hole physics directly. Besides gravitational waves, the Event Horizon Telescope will soon be in a situation to probe the horizon of Sgr $A^*$, the black hole at the center of our galaxy, by studying its shadow directly. Recent progress in observational astronomy thus prompts us to try and understand quantum effects on black holes, as we should soon be in a situation able to confront theory with observations. Some of the questions that we can hope to probe with such observations are whether the horizons of black holes are more violent regions of space-time, in contrast to what is expected in general relativity, and what physical effect resolves the curvature singularity at $r=0$. Both questions are crucial if we want to understand the information paradox linked to Hawking radiation \cite{Hawking:1976ra}.

Black holes are amongst the simplest and yet most mysterious objects in our universe. According to the no-hair theorem, they are described by only a few parameters: their masses, angular momenta and charges. Despite this apparent simplicity, they are incredibly challenging as understanding their physics requires merging quantum mechanics and general relativity, which is one of the remaining holy grails of theoretical physics (see e.g. \cite{Calmet:2015fua} for a recent review). 

While we are far away from having a consistent theory of quantum gravity valid at all energy scales, effective field theory (EFT) techniques applied to general relativity lead to a consistent theory of quantum gravity valid up to an energy scale close to the Planck scale \cite{Donoghue:1994dn,Calmet:2013hfa}. Several universal features of quantum gravity can be identified using these EFT methods. The most intriguing of these features is the dynamical non-locality of space-time induced at short distances by quantum effects. 

This non-locality has been shown to have interesting features in cosmology and could indeed avoid the big crunch singularity in a collapsing universe \cite{Donoghue:2014yha}.
Investigating the effects of this non-locality in black hole physics is our main motivation to consider quantum corrections to spherically symmetric solutions in general relativity. In particular, we revisit the issue of quantum corrections to the Schwarzschild black hole solution which have been studied in the past \cite{Duff:1974ud,BjerrumBohr:2002ks}. We identify a complication which has not been realized previously, namely that of how to define a black hole. 

A mathematically consistent way to define a black hole is to define it as a static vacuum solution, i.e., an eternal black hole. If this definition is adopted, we obtain a result that differs from previous investigations. In particular, we show that the classical black hole solution remains a solution in quantum gravity up to quartic order in the non-local curvature expansion. This result is obtained by looking for perturbative solutions around the Schwarzschild black hole metric. Nevertheless, higher curvature operators give rise to non-trivial corrections. The non-local operators that led to singularity avoidances in \cite{Donoghue:2014yha} thus do not affect the singularity in the case of the Schwarzschild black hole solution.

While eternal black holes are mathematically well defined, they may not capture the full physical picture. A real, astrophysical, black hole is the final state of the evolution of a matter distribution, for example of a heavy star, after it has undergone gravitational collapse. This process is certainly not happening in vacuum. This raises the question of how to define a real astrophysical black hole and of how to calculate quantum corrections to its metric.  A non-vanishing energy-momentum tensor could be used to model a collapsing star. At a time when the star has not yet collapsed into a black hole, the star can be described as a static source at a specific time in its evolution. 

Another complication appears due to the non-locality: we are forced to integrate the modified equations of motion over regions of Planck size curvature. One may thus worry about the sensitivity of the EFT to regions of space-time with high curvature and, in particular, to the singularity at $r=0$. This is investigated in a paper by Donoghue and Ibiapina Bevilaqua \cite{DIB}. Clearly the effective field theory breaks down in regions of large curvature, which in turn raises the question if the latter could offer a reliable picture in our case. However, the ultimate ultra-violet physics that dominates regions of large curvature should not affect observables at long distances, i.e. the exterior region of a black hole. Indeed, in an EFT, one expects short distance physics to decouple at low energies \cite{Appelquist:1974tg}. We will be making  this conservative assumption throughout this paper.

The paper is organized as follows. In section 2, we derive the modified non-local equations of motion. We then consider, in section 3, an eternal Schwarzschild black hole and prove that there are no correction up to quadratic order in curvatures while subsequently showing that the leading order correction appears at cubic order. In section 4, we briefly discuss singularities. In section 5, we turn our attention to the quantum correction around a static star. In section 6, we compare our findings with previous work and then conclude in section 7.

\section{Quantum gravity corrections to the equations of motion}

In analogy to chiral perturbation theory, the effective Lagrangian for quantum general relativity is arranged as a derivative expansion that respects general covariance \cite{Donoghue:1994dn,Don2012}. Quantum fluctuations of the metric, or any other massless field\footnote{This is certainly a valid approximation if one is working at energies between the weak scale and the Planck mass.}, result in a non-local effective action organized as an expansion in curvatures \cite{Donoghue:2014yha,Bar1985,Bar1990,Avr1991}. The final outcome is composed of two parts
\begin{align}
\Gamma[g] = \Gamma_{\text{L}}[c_i(\mu)] + \Gamma_{\text{NL}} \ \ ,
\end{align} 
where the first piece comprises the local effective Lagrangian with renormalized constants. The local part of the Lagrangian contains information about the unknown ultra-violet portion of the theory. The second piece is the non-local portion encoding the infra-red effects. To second order in curvatures \cite{Donoghue:2014yha}, we have\footnote{We dropped a total derivative $\Box R$ as it does not affect the equations of motion.} 
	\begin{align}\label{renormlocal}
	\Gamma_{\text{L}}^{\scriptstyle{(2)}}  = \int d^4x \sqrt{g} \left[ \frac{1}{16 \pi G} R + c_1(\mu) R^2 + c_2(\mu) R_{\mu\nu}R^{\mu\nu}+ c_3(\mu) R_{\mu\nu \alpha \beta}R^{\mu\nu \alpha \beta}\right]
	\end{align}
	and
	\begin{align}\label{nonlocalaction}
	\Gamma_{\text{NL}}^{\scriptstyle{(2)}}  =- \int d^4x \sqrt{g} \left[ \alpha R \ln\left(\frac{\Box}{\mu^2}\right)R + \beta R_{\mu\nu} \ln\left(\frac{\Box}{\mu^2}\right) R^{\mu\nu} + \gamma R_{\mu\nu\alpha\beta} \ln\left(\frac{\Box}{\mu^2}\right)R^{\mu\nu\alpha\beta} \right] \ \ .
	\end{align}
	The various coefficients are given in \cite{Donoghue:2014yha}. In fact, we can invoke the Gauss-Bonnet theorem to express the local action in terms of two independent invariants. We choose to eliminate the Riemann tensor in Eq. (\ref{renormlocal}) which changes the coefficients to
	\begin{align}
	\bar{c}_1 = c_1 - c_3 , \quad \bar{c}_2 = c_2 + 4 c_3, \quad \bar{c}_3=0 \ \ .
	\end{align}
	
	The resulting equations of motion are
	\begin{align}\label{intensoreom}
	G_{\mu\nu} + H_{\mu\nu} +H^{\text{q}}_{\mu\nu} = 0 \ \ ,
	\end{align}
	where $G_{\mu\nu}$ is the Einstein tensor, $H_{\mu\nu}$ and $H^{\text{q}}_{\mu\nu}$ comprise respectively the local and non-local parts of the quantum correction to the field equations. The local contribution is given by
	\begin{align}\label{Hlocal}
	\nonumber
	H_{\mu\nu} &= \bar{c}_1(\mu) \left( 2 R R_{\mu\nu} -\frac{1}{2} g_{\mu\nu} R^2- 2 g_{\mu\nu}\Box R + 2 \nabla_\mu \nabla_\nu R\right)\\
	&+ \bar{c}_2(\mu) \left(\nabla_\alpha \nabla_\mu R^\alpha_\nu + \nabla_\alpha \nabla_\nu R^\alpha_\mu - \Box R_{\mu\nu} -\frac12 g_{\mu\nu} \Box R -\frac12 g_{\mu\nu} R_{\alpha\beta}R^{\alpha\beta} + 2 R_\mu^\alpha R_{\nu\alpha} \right) \ \ .
	\end{align}
Let us now consider the remaining tensor $H^{\text{q}}_{\mu\nu}$  which is the result of varying the non-local action. A consistent method to vary the logarithm has been constructed in \cite{Donoghue:2015nba} and have been shown to contribute to the equation of motion terms cubic in curvatures. The result is indeed quite complicated, however, we do not consider such terms as our aim is to study the theory only up to quadratic order\footnote{Note, nevertheless, that these contributions to the equations of motion are of the same order in derivatives as the ones considered here.}. The non-local contribution to the equations of motion reads
\begin{align}\label{Hnonlocal}
\nonumber
H^{\text{q}}_{\mu\nu}&= \frac12 g_{\mu\nu}  \left[\alpha R \ln\left(\frac{\Box}{\mu^2}\right)R + \beta R_{\alpha \beta} 		\ln \left(\frac{\Box}{\mu^2}\right)R^{\alpha \beta} + \gamma R_{\alpha \beta \sigma \tau} \ln\left(\frac{\Box}				{\mu^2}\right)R^{\alpha \beta \sigma \tau} \right]\\
	&-\left[ 2 \alpha \frac{\delta R}{\delta g^{\mu\nu}} \ln\left(\frac{\Box}{\mu^2}\right)R + 2 \beta  \frac{\delta R_\alpha^\beta}{\delta g^{\mu\nu}} \ln\left(\frac{\Box}{\mu^2}\right) R^\alpha_\beta +  \gamma \frac{\delta R_{	\beta \alpha  \sigma \tau}}{\delta g^{\mu\nu}} \ln\left(\frac{\Box}{\mu^2}\right) R^{ \beta \alpha  \sigma \tau}  \right . \\  \nonumber & \left . 
	+ \gamma \frac{\delta R^{\beta \alpha  \sigma \tau}}{\delta g^{\mu\nu}} \ln\left(\frac{\Box}{\mu^2}\right) R_{\beta \alpha  \sigma \tau} \right] \ \ .
	\end{align}
 Note that there is no way to place the indices on the Riemann tensor such that the last piece is manifestly symmetric. Hence, we choose to vary each tensor separately. The variation of both the Ricci tensor and Ricci scalar is simple but we choose not to display since these terms do not effect our analysis in the next section. The variation of the last two terms in Eq. (\ref{Hnonlocal}) yields
\begin{align}\label{varriemann}
H^{\text{q}}_{\mu\nu}\subset \left(\nabla^\alpha \nabla^\tau + \nabla^\tau \nabla^\alpha\right) \ln\left(\frac{\Box}{\mu^2}\right)  R_{\alpha\nu\mu\tau} + \left(\delta^\alpha_\mu R_{\nu}^{~\beta\sigma\tau} + \delta_\nu^\alpha R_{\mu}^{~\beta\sigma\tau} \right)\ln\left(\frac{\Box}{\mu^2}\right) R_{\alpha\beta\sigma\tau}  \ \ .
	\end{align}
We are now in a position to investigate corrections to the Schwarzschild solution.

\section{Absence of perturbative correction to Schwarzschild black hole}

In this section we look for spherically symmetric black holes in the vacuum of the theory including the full set of quantum corrections up to quadratic order in curvatures. Indeed, this is quite complicated to do in full generality since the equations of motion are integro-differential. Analytic solutions are almost impossible to find and ultimately one must resort to numerical methods. Here, we instead look for linearized solutions around the Schwarzschild black hole solution. This, on one hand, gives us analytic handle on the problem and it conforms to the expectation of the effective theory framework in the sense that quantum-induced corrections should be  {\em small} compared to the classical solution.
	
Precisely, we write the metric as follows
\begin{align}
g_{\mu\nu} = g_{\mu\nu}^{\text{Sch.}} + g^{\text{q}}_{\mu\nu} \ \ ,
\end{align}
where $g^{\text{q}}$ represents the quantum correction to Schwarzschild solution. Linearizing Eq. (\ref{intensoreom}) around $g_{\mu\nu}^{\text{Sch.}}$ one finds
\begin{align}\label{eomlinear}
 G^{\text{L}}_{\mu\nu} \left[g^{\text{q}}\right] + H_{\mu\nu} \left[g^{\text{Sch.}}\right] + H^{\text{q}}_{\mu\nu}\left[g^{\text{Sch.}}\right] = 0 \ \ ,
\end{align}
where the linearized Einstein tensor reads
\begin{align}
G^{\text{L}}_{\mu\nu} = \Box g^{\text{q}}_{\mu\nu} - g_{\mu\nu} \Box g^{\text{q}} + \nabla_\mu \nabla_\nu g^{\text{q}} + 2 R^{\alpha~\beta}_{~\mu~\nu} g^{\text{q}}_{\alpha\beta} - \nabla_\mu \nabla^\beta g^{\text{q}}_{\nu\beta} - \nabla_\nu \nabla^\beta g^{\text{q}}_{\mu\beta} + g_{\mu\nu} \nabla^\alpha \nabla^\beta g^{\text{q}}_{\alpha\beta} \ \ .
	 \end{align}
Clearly, all the terms in Eq. (\ref{Hlocal}) vanish identically, i.e. $H_{\mu\nu}\left[g^{\text{Sch.}}\right] = 0$. This is true as well for all terms proportional to the Ricci tensor or Ricci scalar in $H^{\text{q}}_{\mu\nu}$. Moreover, using the Bianchi identity one can easily show that
\begin{align}
 \left(\nabla^\alpha \nabla^\tau + \nabla^\tau \nabla^\alpha\right) R_{\alpha\nu\mu\tau} = 0
\end{align}
for the Schwarzschild solution\footnote{Here, we freely commuted the derivative operators past the logarithm in Eq. (\ref{varriemann}). This is admissible because the logarithm is precisely a bi-local function that could only depend on the geodesic distance between any two points.}. At this stage one has to specify what the operator $\ln (\Box)$ means in curved space. It has been shown in \cite{Donoghue:2014yha} that the latter is in fact a non-local (bi-local) tensor. Formally, a convenient way to resolve a non-local form factor is as follows \cite{Donoghue:2014yha}
\begin{align}
L(x,x^\prime;\mu) \equiv \ln\left(\frac{\Box}{\mu^2}\right) \frac{\delta(x-x^\prime)}{\sqrt{g}^{1/2}\sqrt{g}^{1/2}} \ \ .
\end{align}
 The quantum corrections to the equations of motion then take the form
\begin{align}\label{tensoreom}
G^{\text{L}}_{\mu\nu} = \left(\delta^\alpha_\mu R_{\nu}^{~\beta\sigma\tau} + \delta_\nu^\alpha R_{\mu}^{~\beta\sigma\tau} - \frac12 g_{\mu\nu} R^{\alpha\beta\sigma\tau} \right) \int d^4x^\prime \sqrt{g}\, L(x,x^\prime;\mu) R_{\alpha\beta\sigma\tau}  \ \ ,
\end{align}
where the right-hand side is to be evaluated using the Schwarzschild metric.
One may worry that the non-local structure forces us to work with coordinates which are regular across the horizon since the integral probes all of space-time. One simple choice is the Eddington-Finkelstein coordinates
\begin{align}\label{EFmetric}
 g_{\mu\nu}^{\text{Sch.}} dx^\mu dx^\nu = \left(1-\frac{2GM}{r}\right) dv^2 - 2 dv dr - r^2 d\Omega^2 \ \ .
\end{align}
We shall nevertheless see that our results do not depend (as expected because of general covariance) on the choice of coordinates and one can work as well with the standard Schwarzschild coordinates.
	 
Indeed, the function $L(x,x^\prime;\mu)$ is very complicated to write down for any particular background space-time. Nevertheless, looking at the metric in Eq. (\ref{EFmetric}) one easily deduces that
 \begin{align}
	 L(x,x^\prime;\mu) = L^{\text{flat}}(\vec{x}-\vec{x}^\prime) + \mathcal{O}(\partial g) \ \ .
	\end{align}
	 Working in flat space-time we find
\begin{eqnarray}
L^{\text{flat}}(\vec x- \vec x^{ \prime})&=&\log \frac{\Box}{\mu^2}=\int_0^\infty ds \left ( \frac{1}{\mu^2+s} -\frac{1}{\Box+s} \right) \\
                   &=&\int_0^\infty ds \left ( \frac{1}{\mu^2+s} - G(\vec x- \vec x^{ \prime}, \sqrt{s})\right)
\end{eqnarray}
where
\begin{eqnarray}
(\nabla^2 + k^2) G(\vec x-\vec x^\prime, k)=\delta(\vec x-\vec x^\prime)
\end{eqnarray}
with
\begin{eqnarray}
G(\vec x- \vec x^{ \prime}, k)=- \frac{e^{i k |\vec x-\vec x^{ \prime}|}}{4 \pi |\vec x-\vec x^{ \prime}|} \ \ .
\end{eqnarray}
After some massaging, we obtain an expression for $L^{\text{flat}}(\vec x- \vec x^{ \prime})$:
	\begin{align}\label{logfinal}
	L^{\text{flat}}(\vec x- \vec x^{ \prime}) = -\frac{1}{2\pi} \lim_{\epsilon \rightarrow 0} \left[\mathcal{P}\left(\frac{1}{|\vec{x}-		\vec{x}^\prime|^3}\right) + 4\pi (\ln(\mu\epsilon)+\gamma_E-1) \delta^{(3)}(\vec{x}-\vec{x}^\prime)\right] \ \ .
\end{align}
where the principle value $\mathcal{P}$ is defined by
\begin{eqnarray}
\mathcal{P}
\left(\frac{1}{f}\right)[g] \equiv \lim_{\epsilon \to 0} =\sum_{x_i |f(x_i)=0} \left [ \int_{-\infty}^{x_i -\epsilon} dx \frac{g(x)}{f(x)} +
\int^{\infty}_{x_i +\epsilon} dx  \frac{g(x)}{f(x)} \right ] \ \ .
\end{eqnarray}

Armed with this expression, we can now show that the right-hand side of Eq. (\ref{tensoreom}) vanishes when applied on the solution. With appropriate placement of indices on the Riemann tensor, all integrals which appear in Eq. (\ref{tensoreom}) have the following form
	\begin{align}\label{central}
	\int d^3x^\prime \sqrt{g} L^{\text{flat}}(\vec x- \vec x^{ \prime}) \frac{1}{(r^\prime)^\text{n}} = - \frac{2}{r^\text{n}} \ln (\mu r),
	\end{align}
	This is a crucial point in the derivation: there are no constant terms accompanying $\ln (\mu r)$. One can go ahead and evaluate the right-hand side of Eq. (\ref{tensoreom}) for any pair $(\mu,\nu)$ to find that it vanishes identically. This is simple to understand given that $\ln \mu$ comprises a local contribution and must vanish when the right-hand side is evaluated on the Schwarzschild solution. It is remarkable we obtain a finite result despite the integral in Eq. (\ref{central}) extends all the way down to $r=0$. This is due to the properties of the non-local function and is a sign of the self-consistency of the EFT.  
	
	Equivalently, one can also calculate the contribution of the term $R_{\mu\nu\alpha\beta} \ln(\Box/\mu_\alpha ^2)R^{\mu\nu\alpha\beta}$ to the equations of motion by varying this term with respect to the metric functions\footnote{Here, we are using the standard static line element $ds^2 = A(r)dt^2 - B(r) dr^2 - r^2 d\Omega_{2}$.} $A(r)$ and $B(r)=A(r)^{-1}$. Using the usual Schwarzschild metric we find that this term gives the following contribution to the field equations 
\begin{eqnarray}
-\frac{2 M G_N}{\pi} \mathcal{P} \int dr_1 \frac{(r-r_1) (|r-r_1|-|r+r_1|)(48 M G_N r^3+6 M G_N r^2 r_1- 2 M G_N r_1^3-24 r^4) }{r^3 r_1^4(2 M G_N-r)|r-r_1||r+r_1|}
\end{eqnarray} 
which vanishes since the integrant is not singular in the limit $r_1\to r$, i.e. the principle value integral is equal to zero. This also shows that our result is coordinate independent as expected.
	
While this shows that there are no corrections at quartic order in curvature, which is in sharp contrast with previous results, there will be corrections at higher order for example higher dimensional operators such as $c_6 R^{\mu\nu}_{\ \ \alpha\sigma}R^{\alpha\sigma}_{\ \ \delta\gamma}R^{\delta\gamma}_{\ \ \mu\nu}$ will lead to quantum corrections of the Schwarzschild solution. We find the following equation of motion:
 \begin{eqnarray}
\frac{\bar M_P^2 r (r h^{\prime\prime}(r)+2 h^\prime(r))}{2 (r h^\prime(r) + h(r)+1)^{3/2}}+c_6 24 M^2 G_N^2\frac{98 M G_N-45 r}{r^6(r-2 M G_N)}=0
\end{eqnarray}
where we are doing perturbation around the standard Schwarzschild solution $A(r)=1-\frac{2 M G}{r}+h(r)$. Far away from the hole, we find\footnote{Note that there are several local as well as non-local operators at cubic order that would contribute similarly.}
 \begin{eqnarray}
h(r)=c_6 \frac{576 \pi G_N^3 M^2}{r^6}  \ \ .
\end{eqnarray}
This simply demonstrates that the Schwarzschild solution is not a solution of the field equations when higher dimensional operators of dimensions $d \geq 6$ are included.

\section{Singularity avoidance?}	
An immediate consequence of our result is that the singularity avoidance observed in \cite{Donoghue:2014yha} is non-universal as the very same operators do not cure the curvature singularity of an eternal Schwarzschild black hole. However, it is important to keep in mind that these results are obtained in perturbation theory. 

We actually have indications that perturbation theory will break down below the reduced Planck mass.  The non-local operators (\ref{nonlocalaction}) lead to a modification of the propagator for the graviton given by 
\begin{eqnarray}
P^{\mu\nu\rho\sigma}=\frac{L^{\mu\nu\rho\sigma}}{2 p^2 \left (1-\frac{1}{120 \pi} G_N N p^2 \log\left (\frac{-p^2}{\mu^2}\right)\right)}
\end{eqnarray}
where $N=N_s +3 N_f +12 N_V$ with $N_s$, $N_f$ and $N_V$ denoting respectively  the number of scalar, spinor and vector fields in the theory. It has been pointed out \cite{Calmet:2014gya} that this propagator has a complex pole
besides the usual massless pole
\begin{eqnarray}
q^2_{1}&=0,\\ \nonumber
q^2_{2}&=& \frac{1}{G_N N} \frac{120 \pi}{ W\left (\frac{-120 \pi M_P^2}{\mu^2 N} \right)} \ \ , \\ \nonumber
q^2_{3}&=&(q^2_{2})^* \ \ ,
\end{eqnarray}
where $W(x)$ is the Lambert W-function.  The complex pole corresponds to new states with a mass and a width given by $q_0^2=(m-i\Gamma/2)^2$. The position of this complex pole, i.e., the new scale $\Lambda_{\text{NP}}=\text{Re}\, q_0$, which depends on the number of particles in the theory, determines the energy scale at which strong quantum gravitational effects are expected to become relevant. At energies of the order of $\Lambda_{\text{NP}}$, we cannot truncate the effective action at quadratic order and need to consider the full effective action.  It is very likely that this non-perturbative phenomenon is responsible for singularity avoidance.  The physical picture that emerges is that the complex states, having a width, are extended objects which will be created when short distances, i.e., singularities, are probed. These extended objects will screen the singularity from being probed during physical processes and thus lead to singularity avoidance.

	\section{Corrections to the gravitational field of a static source}

It is well known that Birkhoff's theorem\footnote{Birkhoff's theorem states that any spherically symmetric solution of the vacuum field equations must be static and asymptotically flat. This theorem implies that the exterior solution must be given by the Schwarzschild metric.} holds only true in classical Einstein's gravity; see for example \cite{Stelle:1977ry,Lu:2015psa}. It is quite interesting to understand the effect of the quantum induced non-locality on the field of a static spherically symmetric object such as a star. One should expect a non-trivial correction to arise in this case albeit the fact that the Schwarzschild metric remains a solution to the vacuum non-local equations of motion. This is a violation of Birkhoff's theorem even at the perturbative level as we show next. 

We aim for a simplified treatment and thus only the Ricci scalar term in Eq. (\ref{nonlocalaction}) is considered. We use a perturbative approach which is similar to that of section 3. The solution to Einstein equation for a constant density star is readily obtained in closed form, see for example \cite{Wald:1984rg}. Outside the star, the non-locality introduces a non-trivial contribution to Eq. (\ref{eomlinear})
\begin{align}
G^{\text{L}}_{\mu\nu} = \alpha (16 \pi G_N)^2 (\nabla_\mu \nabla_\nu - g_{\mu\nu} \Box) \int_{\text{S}} d^4x^\prime \sqrt{g} \,L(x,x^\prime;\mu)\, T \ \ ,
\end{align}
where the integral extends only over the source region, $T = \rho_0 - 3 P$ is the trace of the energy-momentum tensor, $\rho_0$ is the mass density and $P$ is the pressure. Note that here the perturbative treatment is insensitive to the ultra-violet, the local pieces in Eq. (\ref{renormlocal}) drop out. Both the pressure and metric functions are known in the interior of the star \cite{Wald:1984rg}, for example,
\begin{align}
P(r) = \rho_0 \frac{(1-2GM r^2/R_S^3)^{1/2} - (1 - 2GM /R_S)^{1/2}}{(1 - 2GM /R_S)^{1/2} - 3 (1-2GM r^2/R_S^3)^{1/2}} \ \  ,
\end{align}
 where $R_S$ is the radius of the star. To analyze the field far away from the source, it is enough to expand the right-hand side in powers of $G_N$. To lowest order, we have\footnote{Note that we are working with flat-space derivatives in spherical coordinates.}
\begin{align}
G^{\text{L}}_{\mu\nu} = (2 \pi \rho_0 \alpha) (16 \pi G_N)^2 \left(\partial_\mu \partial_\nu - \eta_{\mu\nu} \partial^2\right) \left[\frac{R_S}{r} + \ln\left( \frac{r+R_S}{r-R_S} \right) \right]
\end{align}
where $R_S$ is the radius of the star and the pressure drops out since it is $\mathcal{O}(G)$. We fix the gauge by looking for spherically symmetric perturbations 
\begin{align}
g^{\text{q}}_{\theta\theta} = g^{\text{q}}_{\phi\phi} = 0 \ \ .
\end{align}
Far away from the source, we find the leading correction
\begin{align}\label{starcorrection}
g^{\text{q}}_{\text{tt}} =  \frac{18 \alpha l^2_\text{P}}{R_S^2} \frac{2 G_N M }{r} , \quad g^{\text{q}}_{\text{rr}} = \frac{12 G_N M \alpha l^2_{\text{P}}}{r^3} \ \ .
\end{align}
Note that it is not possible to recover our previous result for an eternal  Schwarzschild black hole by taking the limit $R_S=0$
as this limit is ill-defined.

\section{Comments on previous results}
	
In Section 3, we established that the Schwarzschild black hole solution furnishes an exact solution to the non-local equations of motion accurate up to quadratic order in curvatures. The expected breakdown of Birkhoff's theorem led us to study the field of a static star and we identified the leading non-trivial quantum correction. In the current section we scrutinize the previous results obtained in \cite{Duff:1974ud,BjerrumBohr:2002ks} and discuss them in light of our findings. We will argue that studying quantum corrections to black holes must be done using the effective action formalism.

We start by linearizing the effective action Eq. (\ref{nonlocalaction}) around flat space 
	\begin{align}\label{lineareom}
	\nonumber
	\partial^2 \left[h_{\mu\nu} - \frac12 \eta_{\mu\nu} h \right] + \kappa^2 &\bigg[ \left(\alpha + \frac{\beta}{4}\right) \eta_{\mu\nu} \partial^4 L^{\text{flat}}(h) - \left( \alpha + \frac{\beta}{2} + \gamma \right) \partial_\mu \partial_\nu \partial^2 L^{\text{flat}}(h) \\
	&+ \left(\frac{\beta}{2} + 2 \gamma \right) \partial^4 L^{\text{flat}}(h_{\mu\nu}) \bigg] = 0 \ \ ,
	\end{align}
	where we used the harmonic gauge, $\kappa^2 = 32\pi G$ and 
	\begin{align}
	L^{\text{flat}}(h_{\mu\nu}) \equiv \int d^4x^\prime \, L^{\text{flat}}(x-x^\prime)h_{\mu\nu}(x^\prime) \ \ .
	\end{align}
	We ignore the local action in Eq. (\ref{renormlocal}) as we only need to track the $\ln \mu^2$ piece to determine their effect. To follow the treatment in \cite{Duff:1974ud,BjerrumBohr:2002ks}, one looks for solutions in the following form
	\begin{align}
	h_{\mu\nu} = h^{(0)}_{\mu\nu}(\vec{x}) + h^{(1)}_{\mu\nu}(\vec{x}), \quad h_{00}^{(0)} =- \frac{2GM}{r}, \quad h_{ij}^{(0)} = -\frac{2GM}{r} \delta_{ij} \ \ .
	\end{align}
	Using the graviton values for ($\alpha,\beta,\gamma$) \cite{Donoghue:2014yha}, one finds
	\begin{align}\label{previousres}
	h^{(1)}_{00} = -\frac{43G^2 M}{15 \pi r^3}, \quad h^{(1)}_{ij} = - \frac{G^2 M}{15 \pi r^3} \left[43\, \delta_{ij} - 176\, x_i x_j - 44\, P_{ij} \ln (\mu r)\right] 
	\end{align} 
	which precisely agrees with \cite{Duff:1974ud,BjerrumBohr:2002ks}. One could easily transform the above results to spherical polar coordinates but we do not display this here.
	
	We are interested to compare this result  with ours that we obtained using the full non-linear action. 
	
	\begin{itemize}
	\item The result is proportional to the combination $\ln (\mu r)$. The problem is not that the metric depends on the renormalization scale\footnote{The authors of \cite{Duff:1974ud,BjerrumBohr:2002ks} argued for dropping the $\ln \mu$ piece as they considered it problematic to have explicit dependance on an unphysical scale. Working with the effective action, we know that this causes no concern. Dimensional transmutation of the Wilson coefficients would turn $\mu$ into a physical ultra-violet scale.}, but rather the mere dependance on the term $\ln (\mu r)$. Our analysis in the previous sections confirms that this dependance must be superfluous. It is precisely the curvature expansion that remedies this problem. 
	
	\item The Newtonian $1/r$ behavior of the lowest order solution, i.e. $h^{(0)}_{\mu\nu}(\vec{x})$, is strictly valid for an idealized point mass. Indeed and due to the uniqueness of Schwarzschild metric in Einstein gravity, a black hole {\it looks} like a point mass to an asymptotic observer. This is not true anymore once we deviate from classical general relativity and hence one has to be careful when interpreting the meaning of quantum corrections.
	
	\item The other pieces, i.e. the $1/r^3$ power law in Eq. (\ref{previousres}), resemble the correction to the $g^{\text{q}}_{\text{rr}}$ component of a static star found in Eq. (\ref{starcorrection}). Note, however, that it is impossible for the linearized analysis to pick the power-law behavior in $g^{\text{q}}_{\text{tt}}$ as the source is treated as a point mass. 	
	\end{itemize}
	
	We conclude that the previous results of \cite{Duff:1974ud,BjerrumBohr:2002ks} should be re-interpreted with great care. They capture {\it some} of the quantum corrections around a massive object such as a star. On the other hand, as shown previously, the Schwarzschild black hole solution remains a solution including non-local quantum corrections up to quadratic order in curvatures.

\section{Conclusions}

Effective field theory techniques applied to general relativity lead to a consistent theory of quantum gravity up to energy scales close to the Planck scale. A key feature of this  effective theory approach is the appearance of non-local effects which become appreciable close to the Planck scale and could, in principle, resolve singularities. Using the EFT approach, we have computed quantum corrections to spherically symmetric solutions of Einstein gravity and focussed in particular on the Schwarzschild black hole solution. We worked to quadratic order in curvatures simultaneously taking local and non-local corrections into account. We searched for solutions perturbatively close to that of classical general relativity, and found that an eternal Schwarzschild black hole remains a solution and receives no quantum corrections up to this order in the curvature expansion. This is in contrast to previous works. We have identified the reason for the discrepancy which can be traced back to the fact that previous results considered a linearized theory. We have also shown that while there are no quantum gravitational corrections to an eternal black hole at quadratic order in curvature, corrections will appear at higher order.  In contrast, and due to the breakdown of Birkhoff's theorem, the field of a massive star receives corrections which are fully determined by the effective theory. 

We have shown that while the non-locality forces us to integrate the equations of motion over regions of space-time with large curvature, we obtain finite results which is a sign of the self-consistency of the effective field theory. Indeed in an effective field theory, one expects a decoupling of scales and long distance corrections should be calculable without requiring any knowledge of the full theory (i.e. ultra-violet physics or singularity in our case). We have also shown that the non-local terms do not soften the singularity in the case of black holes, at least within a perturbative framework.

Our findings emphasize the need to be very careful when discussing quantum corrections to black holes which need to be defined carefully. While, from a mathematical point of view, an eternal black hole is a static vacuum solution, astrophysical black holes are not. They are surrounded by matter and are themselves the result of the gravitational collapse of matter. Calculating quantum gravitational corrections to real astrophysical black holes is thus a fantastically difficult task which cannot be done easily analytically. This investigation requires us to study a dynamical process where a matter distribution, e.g., a star, collapses to form a black hole and to follow quantum effects throughout the process. Our work represents a first step in that direction. We have found that an observer far away from a star experiences a correction to Newton's law that depends on the size of the star. Long after the star has collapsed, the far field behavior of the remaining object should approach that of an eternal black hole. At this stage of the evolution, the observer would find only cubic order in curvature corrections to Newton's law.

\bigskip{}

\noindent{\it Acknowledgments:}
We would like to thank John Donoghue and Leandro Ibiapina Bevilaqua for numerous discussions.
This work is supported in part by the Science and Technology Facilities Council (grant number  ST/L000504/1).


\baselineskip=1.6pt 

\begin{thebibliography}{10}

\bibitem{Hawking:1976ra} 
  S.~W.~Hawking,
  ``Breakdown of Predictability in Gravitational Collapse,''
  Phys.\ Rev.\ D {\bf 14}, 2460 (1976).

\bibitem{Calmet:2015fua} 
  X.~Calmet (ed.),
  ``Quantum aspects of black holes,''
  Fundam.\ Theor.\ Phys.\  {\bf 178} (2015).
  doi:10.1007/978-3-319-10852-0.
  
\bibitem{Donoghue:1994dn} 
  J.~F.~Donoghue,
  ``General relativity as an effective field theory: The leading quantum corrections,''
  Phys.\ Rev.\ D {\bf 50}, 3874 (1994).

\bibitem{Calmet:2013hfa} 
  X.~Calmet,
  Int.\ J.\ Mod.\ Phys.\ D {\bf 22}, 1342014 (2013)
  doi:10.1142/S0218271813420145
  [arXiv:1308.6155 [gr-qc]].
  
\bibitem{Donoghue:2014yha} 
  J.~F.~Donoghue and B.~K.~El-Menoufi,
  ``Nonlocal quantum effects in cosmology: Quantum memory, nonlocal FLRW equations, and singularity avoidance,''
  Phys.\ Rev.\ D {\bf 89}, no. 10, 104062 (2014).

\bibitem{Duff:1974ud} 
  M.~J.~Duff,
  ``Quantum corrections to the schwarzschild solution,''
  Phys.\ Rev.\ D {\bf 9}, 1837 (1974).
  
\bibitem{BjerrumBohr:2002ks} 
  N.~E.~J.~Bjerrum-Bohr, J.~F.~Donoghue and B.~R.~Holstein,
  ``Quantum corrections to the Schwarzschild and Kerr metrics,''
  Phys.\ Rev.\ D {\bf 68}, 084005 (2003)
  Erratum: [Phys.\ Rev.\ D {\bf 71}, 069904 (2005)].
  
  \bibitem{DIB}
  J.~Donoghue and L.~Ibiapina Bevilaqua, to appear.
  
  \bibitem{Appelquist:1974tg} 
  T.~Appelquist and J.~Carazzone,
  ``Infrared Singularities and Massive Fields,''
  Phys.\ Rev.\ D {\bf 11}, 2856 (1975).
  
    \bibitem{Don2012} 
  J.~F.~Donoghue,
  ``The effective field theory treatment of quantum gravity,''
  AIP Conf.\ Proc.\  {\bf 1483}, 73 (2012).
  
    \bibitem{Bar1985} 
  A.~O.~Barvinsky and G.~A.~Vilkovisky,
  ``The Generalized Schwinger-Dewitt Technique in Gauge Theories and Quantum Gravity,''
  Phys.\ Rept.\  {\bf 119}, 1 (1985).
  
  \bibitem{Bar1990} 
  A.~O.~Barvinsky and G.~A.~Vilkovisky,
  ``Covariant perturbation theory. 2: Second order in the curvature. General algorithms,''
  Nucl.\ Phys.\ B {\bf 333}, 471 (1990).
  
  \bibitem{Avr1991} 
  I.~G.~Avramidi,
  ``The Covariant Technique for Calculation of One Loop Effective Action,''
  Nucl.\ Phys.\ B {\bf 355}, 712 (1991).

  \bibitem{Donoghue:2015nba} 
  J.~F.~Donoghue and B.~K.~El-Menoufi,
  ``Covariant non-local action for massless QED and the curvature expansion,''
  JHEP {\bf 1510}, 044 (2015)


\bibitem{Stelle:1977ry} 
  K.~S.~Stelle,
  ``Classical Gravity with Higher Derivatives,''
  Gen.\ Rel.\ Grav.\  {\bf 9}, 353 (1978).
  
\bibitem{Lu:2015psa} 
  H.~Lü, A.~Perkins, C.~N.~Pope and K.~S.~Stelle,
  ``Spherically Symmetric Solutions in Higher-Derivative Gravity,''
  Phys.\ Rev.\ D {\bf 92}, no. 12, 124019 (2015).


\bibitem{Calmet:2014gya} 
  X.~Calmet,
  ``The Lightest of Black Holes,''
  Mod.\ Phys.\ Lett.\ A {\bf 29}, no. 38, 1450204 (2014).
  
  \bibitem{Wald:1984rg} 
  R.~M.~Wald,
  ``General Relativity,''
  Chicago, Usa: Univ. Pr. ( 1984) 491p.

 \end{thebibliography}
\end{document}